\documentclass[preprint,amsmath,amssymb,prl,unsortedaddress,superscriptaddress]{revtex4-1}
\usepackage{graphicx}
\usepackage{amsmath}
\usepackage[colorlinks=true,allcolors=blue]{hyperref}
\usepackage[table]{xcolor}
\usepackage{cancel}
\usepackage[symbol]{footmisc}

\makeatletter
\let\@fnsymbol@latex\@fnsymbol
\let\@fnsymbol\@alph
\makeatother

\newcommand*{\abb}[0]{$\text{(MC)}^2$~}

\newcommand*{\fig}[1]{Figure~\ref{fig:#1}}

\newcommand*{\eqn}[1]{Equation~\ref{eqn:#1}}

\newcommand*{\red}[1]{\textcolor{red}{#1}}

\begin{document}

\title{Multi-Cell Monte Carlo Method for Phase Prediction}
\author{C. Niu \footnotemark[1] \footnote[2]{Current address: QuesTek Innovations LLC, Evanston IL 60201, USA}}
\author{Y. Rao  \footnote[1]{C.N and Y. R made equal contributions to this work.} }
\affiliation{Department of Materials Science and Engineering, The Ohio State University, Columbus OH 43210, USA}
\author{W. Windl}
\affiliation{Department of Materials Science and Engineering, The Ohio State University, Columbus OH 43210, USA}
\affiliation{Department of Physics, The Ohio State University, Columbus OH 43210, USA}
\author{M. Ghazisaeidi}
\affiliation{Department of Materials Science and Engineering, The Ohio State University, Columbus OH 43210, USA}
\affiliation{Department of Physics, The Ohio State University, Columbus OH 43210, USA}

\date\today

\begin{abstract}
We propose a Multi-Cell Monte Carlo algorithm, or (MC)$^2$, for predicting stable phases in chemically complex crystalline systems.  Free atomic transfer among cells is achieved via the application of the lever rule, where  an assigned molar ratio virtually controls the percentage of each cell in the overall simulation, making (MC)$^2$ the first successful algorithm for simulating phase coexistence in crystalline solids. During the application of this method, all energies are computed via direct Density Functional Theory calculations. We test the method by successful prediction of the stable phases of known binary systems.  We then apply the method to a quaternary high entropy alloy. The method is particularly robust in predicting stable phases of multi-component systems for which phase diagrams do not exist.
\end{abstract}

\maketitle

Prediction of stable phases of multicomponent systems is a crucial step in understanding thermodynamics of alloys. The increasing availability of first-principles methods, and systematic approaches for predicting possible arrangements of atoms in an alloy, such as the cluster expansion methods\cite{de1994cluster} have paved first inroads to model few-component systems. However, obtaining phase diagrams for ternary and beyond compositions  has remained a challenging and often unfeasible task due to the complexity of the problem. In addition to predicting potential stable phases, an extra challenge is the coexistence of phases and prediction of phase fractions. 
To overcome these restrictions, we present a Multi-Cell Monte Carlo, or \abb, algorithm that is capable of predicting both the coexistence of multiple phases in a chemically complex crystalline system and the composition and structure of the different phases in a single run. To the best of our knowledge, this is the first and only method that can capture the phase boundary from only one initial composition, without the need to interpolate intermediate compositions.

Coexisting supercells in Monte Carlo simulations have previously been used within the Gibbs ensemble Method of Panagiotopoulos \cite{panagiotopoulos1988phase} for simulation of vapor/liquid equilibrium, where atoms are constantly deleted/inserted in the cells. In crystalline solids, deleting/inserting atoms in each cell  creates excessive point defects. Therefore, the Gibbs ensemble MC has not been applied to phase predictions in solids. Recently, we have introduced  the first  multi-cell  MC relaxation  method to solids, which required fixed number of atoms in each cell 
\cite{niu2017multi}. However, the fixed cell sizes there restrict the compositional variations and do not allow for prediction of phase fractions. 

Here, we maintain the idea of a multi-cell MC but eliminate the fixed-size restriction via a modified algorithm.
In the new method, each cell is assigned a molar ratio to virtually control its percentage in the total system. The molar ratios are determined by the compositions of all cells via application of the ``lever rule" so that the total composition of the system is constant.  In a Gibbs ensemble MC simulation, a random atom is transferred from one cell to another under constant-$(\mu,V,T)$ condition. Here, the transfer of random atoms is achieved by changing the molar ratios of cells. Specifically, we randomly change the species of an atom in one or more cells, which we call a \textit{flip} move. A flip move changes the compositions in each cell, which leads to a different set of molar ratios, or different percentages of each cell in the system. As a consequence, this is equivalent to the transfer of a group of random atoms among the cells.  The total energy of each cell is computed with Density Functional Theory (DFT), with relaxation of all degrees of freedom at zero pressure. Note that the flip moves are not arbitrary and need to conserve the total number of each species among all cells.

The lever rule is often introduced in the study of binary phase diagrams \red{\cite{william2006foundations}}. It is a tool to determine the molar or volume ratio of each phase of a binary system at equilibrium at a temperature using only the compositions where the tie line crosses phase boundaries. Assuming two supercells that represent two phases of a binary system, conservation of the initial stoichiometry requires
\begin{align}
\underbrace{\begin{bmatrix} n^a_1 & n^a_2 \\ n-n^a_1 & n-n^a_2\end{bmatrix}}_{\bf{A}}
\underbrace{\begin{bmatrix} x_1 \\ x_2 \end{bmatrix}}_{\bf{X}} =
\underbrace{\begin{bmatrix} n\cdot c^a \\ n\cdot(1-c^a) \end{bmatrix}}_{\bf{B}}
\label{eqn:bin-stoi}
\end{align}
where the superscript $a$ indicates element $a$ while the subscript 1 and 2 indicate the supercells/phases, $n_1^a$ and $n_2^a$ are the number of atoms of element $a$ in supercell 1 and 2, respectively, $n$ is the total number of atoms in the system and $c^a$ is the atomic concentration of element $a$ in the system. 
The molar ratios ($x_1$ and $x_2$) are then obtained from $\bf{X}=\bf{A}^{-1}\cdot \bf{B}$. Note that $\bf{B}$ is a constant vector, given by the initial composition of the alloy and $\bf{A}$ is updated at each MC step.
\eqn{bin-stoi} can be easily generalized to a $m$-component system with $m>2$. 

The compositional fluctuation is achieved by changing the chemical identity of atoms rather than particle insertion/deletion. Specifically, we change the element of a randomly chosen atom to any of the elements in the system with a probability equal to the system concentration of the new element.  We call these moves \textit{flips}. These moves were first introduced  by  Kofke and Glandt to establish the semigrand canonical ensemble~\cite{kofke1988monte}. In our simulation, such  flips can happen locally (i.e. flip a random atom in one random supercell) or globally (i.e. flip a random atom in all supercells at the same time) without causing bias. This particle flip move, combined with the application of the lever rule to multiple cells, mimics the effect of varying cell size and composition in each cell, without the need for particle insertion/deletion in the cells.

Furthermore, in any MC simulation, the  acceptance criterion must satisfy detailed balance within the chosen ensemble. In \abb, two levels of simulation can be envisioned: within individual cells and considering all cells combined. Within each cell $i$, and ignoring all other cells, the total number of atoms $n_i$ is fixed, but the ratio of different species can change. This corresponds to the semigrand canonical ensemble. 

%
%

On the other hand, considering all cells, the total number of atoms $n$ as well as total numbers of each species in the entire simulation is fixed as manifested in matrix $\bf{B}$ of \eqn{bin-stoi}. Therefore, the entire simulation represents a canonical ensemble with acceptance criterion  

\begin{equation}
P=min(1,e^{-\beta[\Delta E_\text{tot}]})
\label{eqn:can}
\end{equation}

\noindent where $\Delta E_\text{tot}$ is the change in energy of all cells combined. The contribution of each cell to the total energy is governed by the molar ratio of the cell, i.e $E_\text{tot}=\sum_i{x_i  E_i}$. Therefore,
$\Delta E_\text{tot}=\sum_i{x^\prime_i  E^\prime_i}-\sum_i{x_i  E_i}$
\noindent where the summation is over all the cells and primed and unprimed variables denote quantities after and before an MC step, respectively.

Accurate calculation of total energies in multicomponent alloys requires first principles calculations. We use the plane-wave-basis density functional theory  code VASP \cite{kresse1999g,kresse1996g}, with 
projector-augmented-wave method pseudopotentials \cite{blochl1994projector,kresse1999ultrasoft}, the Perdew, Burke, and Ernzerhof generalized gradient approximation exchange-correlation functional \cite{perdew1996generalized}, and  a Monkhorst-Pack mesh \cite{monkhorst1976special} for Brillouin zone integration with  Methfessel-Paxton smearing \cite{methfessel1989high}. The cutoff energies used are at least 30\% higher than the default values. Two settings for the DFT calculations are used, \textit{accurate} and \textit{fast}. The accurate setting ensures a total energy convergence of 0.2 meV/atom. The fast settings are used during the MC simulations to make the calculations feasible and have reduced k-point meshes and cutoff energies for MC simulations. Details of settings for individual cases are presented in the Supplementary Material.  Upon completion of a \abb run with \textit{fast} settings, 9 evenly distributed configurations are selected and recalculated with the accurate settings. The difference between energies obtained by the two settings at these points
quantifies the error caused by using less accurate settings during the MC moves and verifies that general trends and observations remain unchanged. All the initial input structures are special quasi-random structures (SQS) \cite{zunger1990special} generated by the Alloy Theoretic Automated Toolkit (ATAT) \cite{van2002alloy} where the structures perfectly mimic those of random alloys in terms of at least first nearest neighbors.

\begin{figure}
\includegraphics[width=3in]{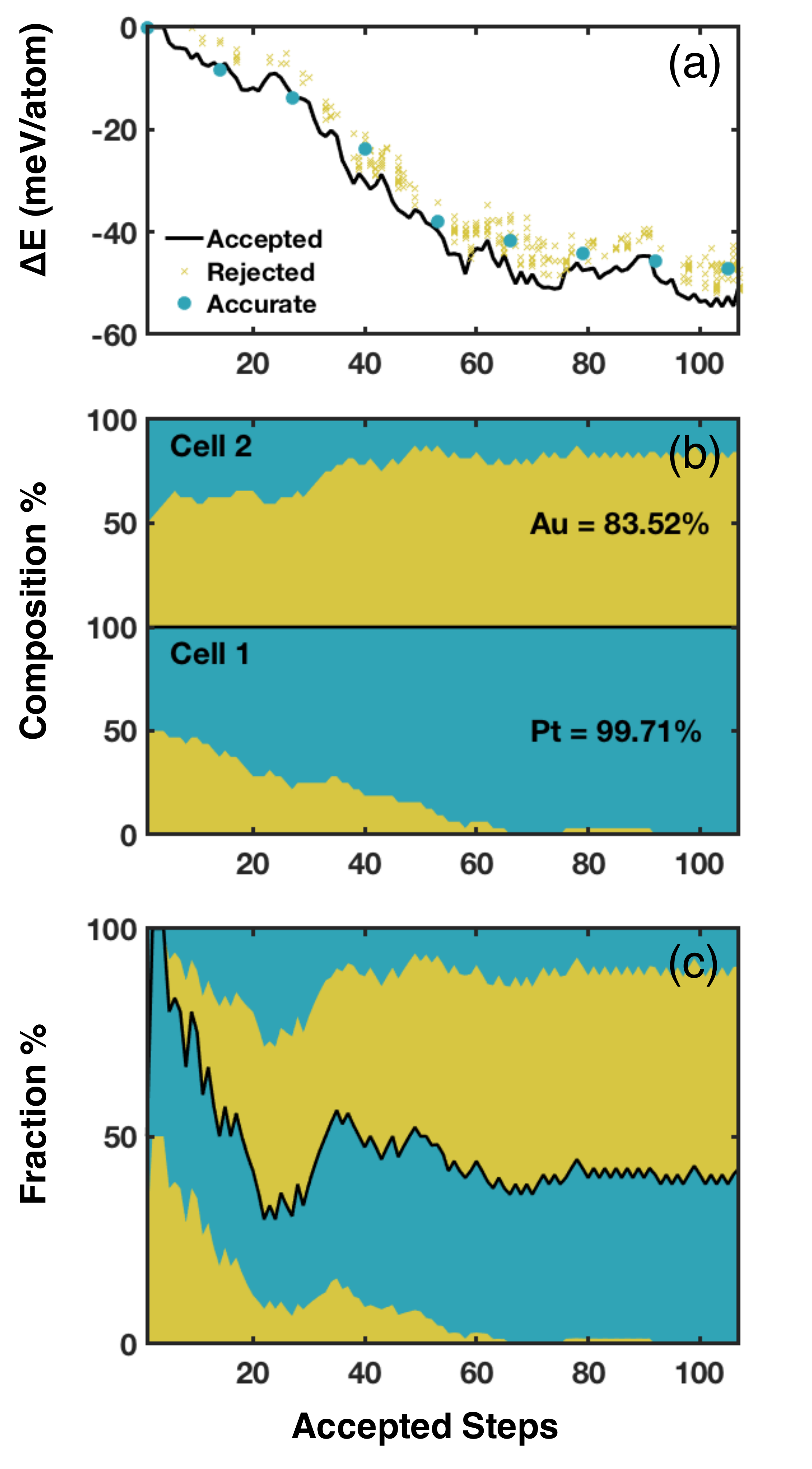}
\caption{Change in energy and composition of AuPt alloy during the applications of \abb at $T=600$~K. (a)  shows the change of energy as a function accepted steps (solid line) and rejected steps ($\times$'s) relative to the initial step. Dots show the energies recalculated with the \textit{accurate} DFT settings. (b) shows the change of composition in each cell. Yellow represents gold and blue represents platinum. The initial composition is equiatomic. Clear phase separation can be seen at the end of run. (c) presents the evolution of molar fraction of each cell.}
\label{fig:AuPt}
\end{figure}

An equiatomic Au-Pt alloy is chosen as our first benchmark, since the Au-Pt phase diagram has a miscibility gap at low temperatures \cite{okamoto1985pt} and AuPt alloys are thus expected to decompose into Pt and Au-enriched phases.

\fig{AuPt} summarizes our prediction of stable phases for this system 
at $T=600$~K. 
\fig{AuPt}(a) shows the change in energy of the system as a function of the accepted  MC steps. Rejected steps are also indicated by $\times$ symbols. The evolution of the composition in each cell can be read from \fig{AuPt}(b), where both cells start with equal numbers of Au and Pt atoms, while after ~100 accepted steps, Cell 1 starts to contain only Pt atoms, while Cell 2 has an Au-rich composition. The actual concentration of each element in the entire system, i.e composition of each cell multiplied by its corresponding molar fraction, is shown in \fig{AuPt}(c), where phase 1, the Pt-only phase in cell 1, forms 40\% of the solid, while phase 2 with $\sim84$\% Au forms $\sim60$\%.   Therefore, \abb not only correctly predicts the separation into a Pt and Au-rich separations, but also yields the relative amount of each phase. \textit{This is the major new finding of this paper, and when performed at various temperature/composition combinations, (MC)$^2$ can predict the phase diagram of the alloy.}

\begin{figure}
\includegraphics[width=3in]{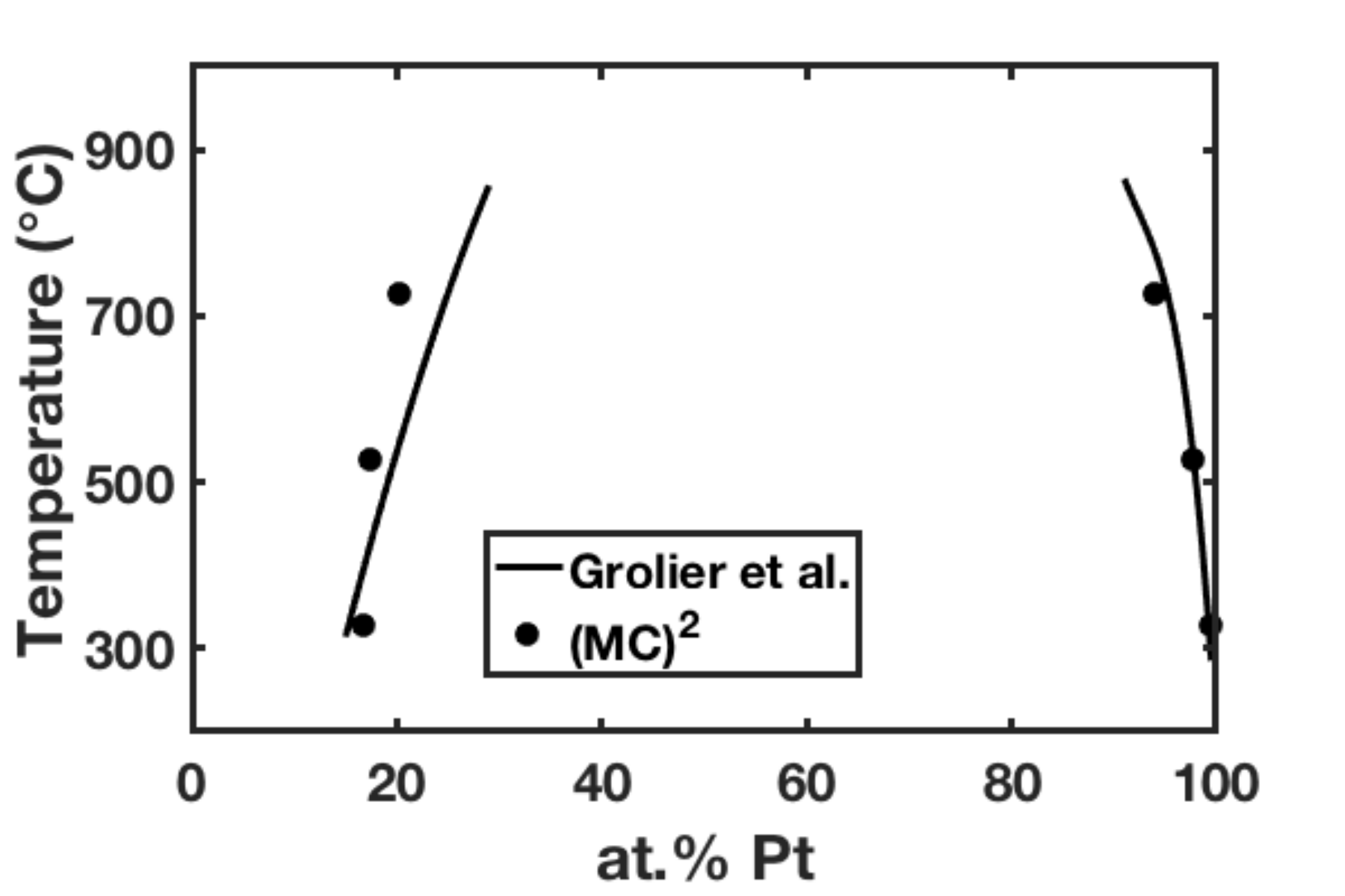}
\caption{\abb Prediction of Au-Pt phase boundaries at 600~K, 800~K, and 1000~K against the  experimental phase diagram by Grolier et al. \cite{grolier2008experimental}.}
\label{fig:diagram}
\end{figure}

\fig{diagram} compares the \abb predictions of Au-Pt phase boundaries at 600~K, 800~K, and 1000~K against the phase diagram by measured by Grolier et al. \cite{grolier2008experimental}. The points at $T=600$~K (323~$^\circ$C) are the average compositions of the last $10\%$ accepted steps of each cell shown in \fig{AuPt}(b). Similarly, other points are obtained by separate \abb runs at $T=800$~K and $T=1000$~K. The 
\begin{figure}[h]
\includegraphics[width=3in]{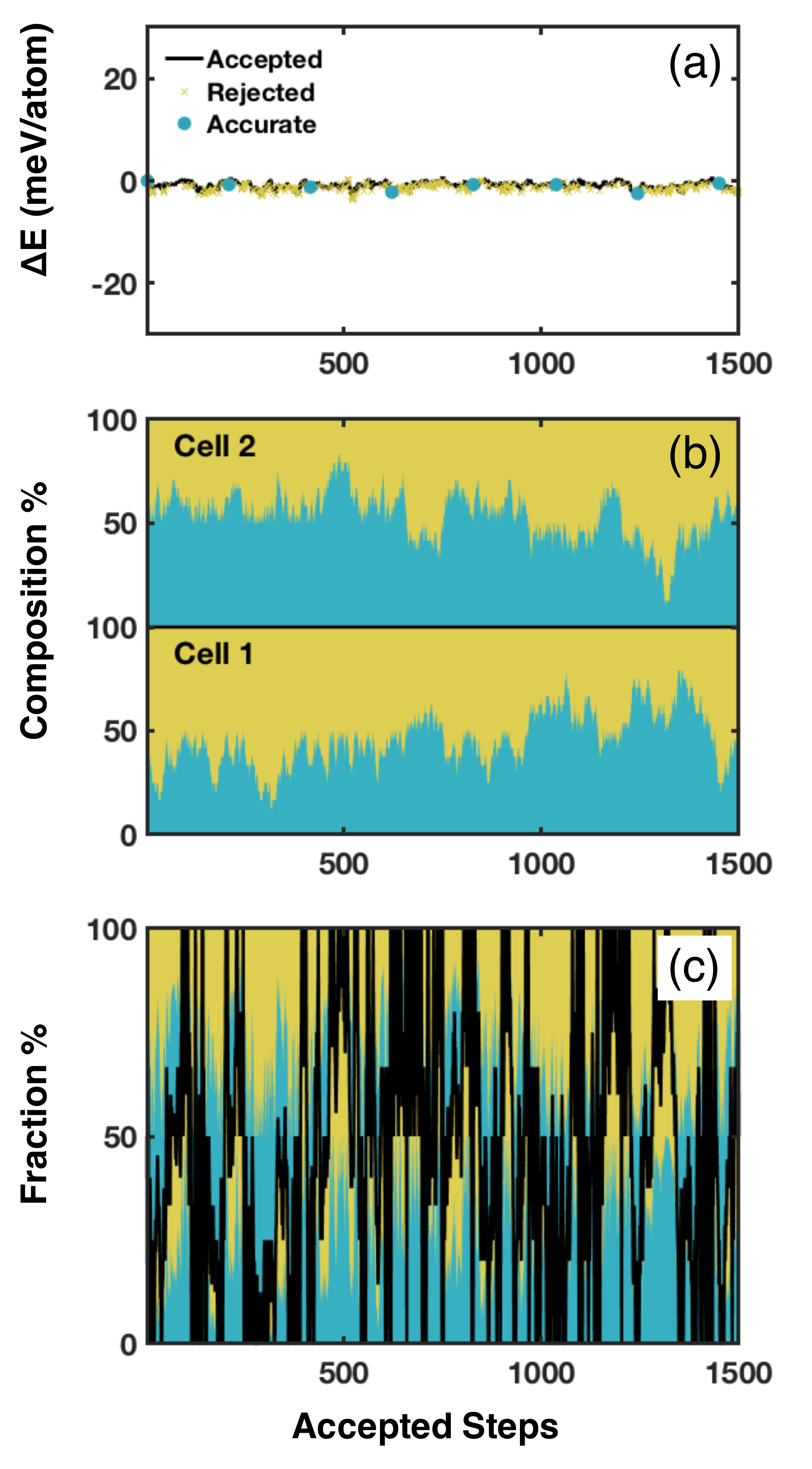}
\caption{\abb  results for (a) change in energy, (b) composition, and (c) molar ratios of an equiatomic HfZr alloy at $T=400$~K. Yellow represents zirconium and blue represents hafnium. 
}
\label{fig:HfZr}
\end{figure}
predicted miscibility gap in this temperature range agrees very well with 
experimental measurements \red{\cite{grolier2008experimental}.}

Next, we test the method with the Hf-Zr binary alloy. \abb successfully predicts the complete solubility of Hf-Zr alloy at $T=400$~K, $T=700$~K and $T=1000$~K. \fig{HfZr} shows the change in energy and composition of an equiatomic Hf-Zr alloy at $T=400$~K. The composition is 50$\%$ Hf,  50$\%$  Zr initially and fluctuates around this value during the entire run without any significant change in energy. The molar fractions also fluctuate around half, confirming that \abb maintains a solid solution phase. The corresponding plots at the other temperatures considered, are similar and are not shown.

The true predictive power of \abb is demonstrated when applied to multicomponent systems, particulary beyond ternary alloys where phase diagrams do not exist. \red{Here, we} study the quaternary, equiatomic, HfNbTaZr high-entropy alloy. This alloy forms a body-centered cubic (bcc) single-phase random solid solution after casting \cite{wu2014refractory}. 

\begin{figure*}
\includegraphics[width=6in]{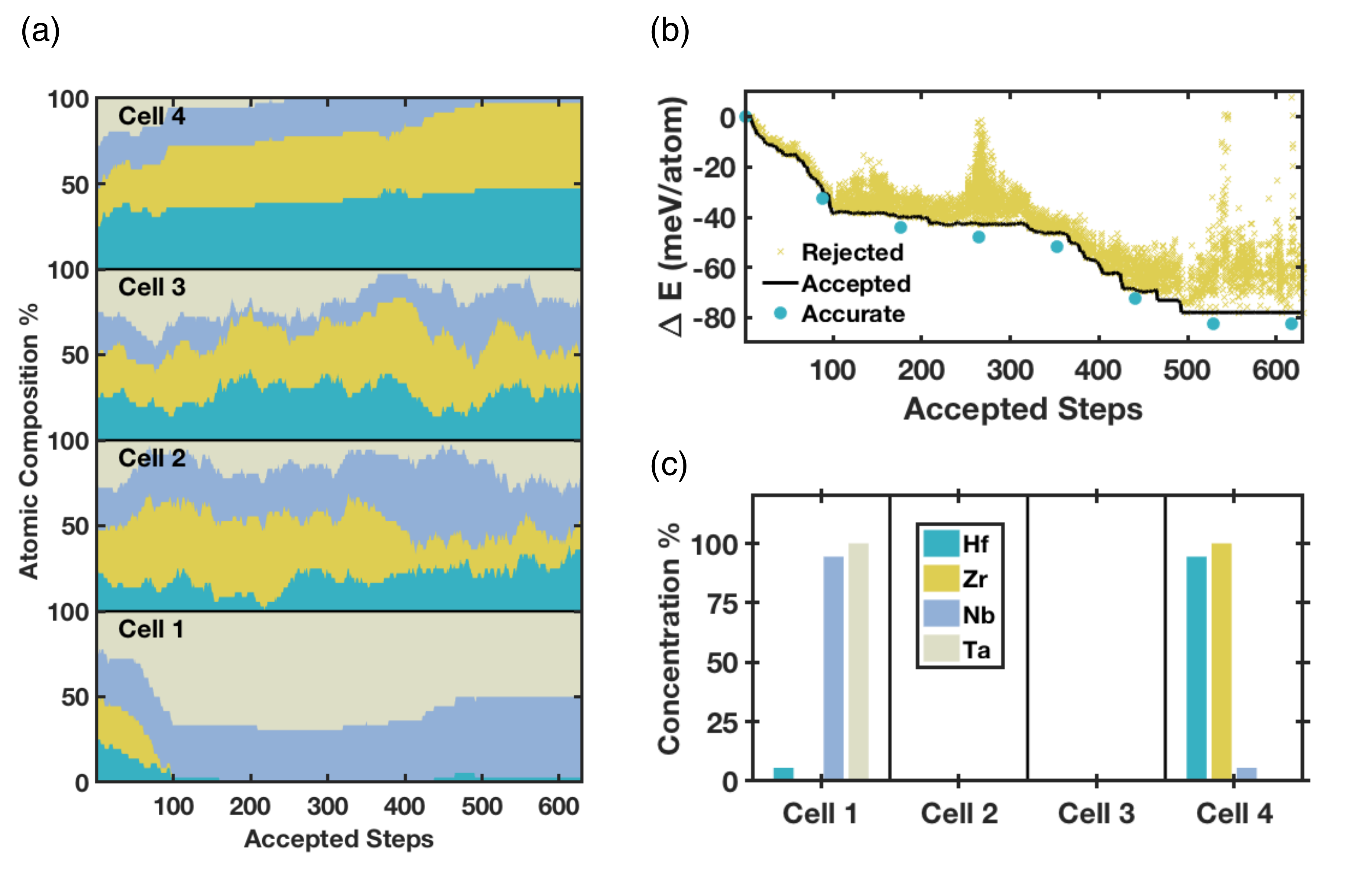}

\caption{Predicted stable phases of the quaternary HfZrTaNb high entropy alloy using 4 cells. (a) shows the change in composition  in each cell. All 4 cells start with the equiatomic random composition of the HEA, while (b) shows the associated energy change as a function of accepted steps. Rejected steps are also shown by the cross symbols. (c) shows the total concentration of each element, (.i.e composition of each cell multiplied by the corresponding molar ratio) at the final step. Cells 2 and 3 have zero molar fractions, implying  phase separation into two phases: a Nb-Ta BCC phase in Cell 1 and a Hf-Zr hcp phase in Cell 4.}
\label{fig:HEA}

\end{figure*}

\fig{HEA} summerizes the predicted stable phases of this quaternary alloy using 4 cells. \fig{HEA}(a) shows the change in composition  in each cell. All 4 cells start with the equiatomic random bcc cells of the HEA. Cells 1 and 4 are gradually enriched by Nb/Ta and Hf/Zr respectively, while the other two cells maintain more or less the equiatomic composition. Note that these are the compositions in each cell during the simulation. For the actual alloy composition, the cell compositions are multiplied by the molar fraction of each cell/phase  shown in \fig{HEA}(c). It is evident that the molar fractions of cells 2 and 3 are zero. In other words these two cells do not represent a phase after the phase separation and must be discarded. 
Therefore, the only remaining phases are the two Nb/Ta-rich, and Hf/Zr-rich ones obtained from cells 1 and 4 respectively. 
Since all degrees of freedom, including cell volume and shapes are optimized by VASP, the Hf/Zr-rich  Cell 4 transforms into a hexagonal close packed (hcp) lattice from the initial bcc lattice. Therefore, \abb predicts the phase separation of the bcc HfZrTaNb HEA into two phases of Nb/Ta (bcc) and Hf/Zr (hcp). These predictions are consistent with those of a recent Atom Probe Tomography study on HfNbTaZr, discovering the formation of a secondary phase enriched in Hf and Zr after sufficient annealing \cite{maiti2016structural}.


In summary, we have introduced the \abb method to predict stable phases and phase fractions in multicomponent alloys.  As we have demonstrated, 
each converged \abb run identifies either a region of miscibility, or the relevant phase boundaries for the simulated temperature and composition.
The algorithm in \abb takes advantage of parallel computations of multiple cells and provides the unique capability of identification of relevant phases and phase boundaries without any prior knowledge of possible phases.
In its current implementation, the MC algorithm takes into account the configurational entropy contributions to the free energy, while other contributions such as vibrational entropy are neglected. This makes phase diagram predictions reliable for temperatures sufficiently below the solidus curve. 

\noindent \textit{Acknowledgments}

Primary funding for this work was provided by the Air Force Office of Scientific Research Grant FA9550-17-1-0168. YR is supported by the National Science Foundation grant DMREF-1534826. WW acknowledges funding from the WastePD, an Energy Frontier Research Center funded by the U.S. Department of Energy, Office of Science, Basic Energy Sciences under Award \# DESC0016584. Computational resources were provided through the Ohio Supercomputer Center.

\bibliography{ref}
\end{document}